\documentclass[preprint,11pt]{aastex}
\usepackage{epsfig}
\usepackage{rotating}      
\tightenlines

\def	\Angstrom	{\,{\rm \AA}}
\def	\ASCA	{{\it ASCA}}
\def	\beq	{\begin{equation}}
\def	\Chandra {{\it Chandra}}
\def	\cm	{\,{\rm cm}}
\def	\eeq	{\end{equation}}
\def	\eV	{\,{\rm eV}}
\def	\fhalo	{f_{\rm halo}}
\def	\FHI	{{\rm FHI}}
\def	\g		{\,{\rm g}}
\def	\gtsim	{\gtrsim}		 
\def	\keV	{\,{\rm keV}}
\def	\kpc	{\,{\rm kpc}}
\def	\ltsim	{\lesssim}		 
\def	\Nhalo	{N_{\rm halo}}
\def	\Nptsrc	{N_{\rm ptsrc}}
\def	\pc	{\,{\rm pc}}
\def	\ROSAT	{{\it ROSAT}}

\def	\tausca	{\tau_{\rm sca}}
\def	\thetah		{\theta_{h}}
\def	\thetahm	{\theta_{h,50}}
\def	\thetahten	{\theta_{h,10}}
\def	\thetahninety	{\theta_{h,90}}
\def	\thetas		{\theta_{s}}
\def	\thetasm	{\theta_{s,50}}
\def	\thetasten	{\theta_{s,10}}
\def	\thetasninety	{\theta_{s,90}}

\newlength{\figwidth}
\newlength{\figwidthland}
\newlength{\figwidthport}
\countdef\decade=200
\advance\decade by \year
\countdef\hours=201
\advance\hours by \time
\divide\hours by 60
\countdef\mins=202
\advance\mins by \hours
\multiply\mins by 60
\multiply\hours by 100
\countdef\miltime=203
\advance\miltime by \hours
\advance\miltime by \time
\advance\miltime by -\mins

\begin{document}

\title{
        \vspace*{-3.0em}
        {\normalsize\rm To appear in {\it The Astrophysical Journal}}\\ 
        \vspace*{1.0em}
	Scattering by Interstellar Dust Grains. II. X-Rays
	}

\author{B.T. Draine			
	}
\affil{Princeton University Observatory, Peyton Hall, Princeton,
NJ 08544; \\
{\tt draine@astro.princeton.edu}}

\begin{abstract}
Scattering and absorption of X-rays by interstellar dust
is calculated
for a model consisting of carbonaceous grains
and amorphous silicate grains.
The calculations employ realistic dielectric functions with
structure near X-ray absorption edges,
with resulting features in absorption, scattering, and extinction.

Differential scattering cross sections are calculated for 
energies between 0.3 and 10 keV.
The median scattering
angle is given as a function of energy, and
simple but accurate approximations are found for
the X-ray
scattering properties of the dust mixture, as well as for
the angular distribution of the scattered X-ray halo for dust with simple
spatial distributions.
Observational estimates of the X-ray scattering optical depth 
are compared to model predictions.
Observations of X-ray halos 
to test interstellar dust grain models are best carried out using
extragalactic point sources.
\end{abstract}

\keywords{dust, extinction -- 
	polarization -- 
	scattering -- 
	ultraviolet: ISM -- 
	X-rays: ISM}

\section{Introduction
	\label{sec:intro}}

Interstellar grains can absorb and scatter X-rays.  The scattering,
typically through small angles, 
results in a ``halo'' of scattered X-rays within
$\sim1^\circ$ of an X-ray source
(Overbeck 1965; Hayakawa 1970; Martin 1970).
The observed properties of these X-ray halos
provide a test of interstellar
grain models.

The nature of interstellar grains remains uncertain
(see Draine 2003a, and references therein).  
This paper will examine
X-ray scattering and absorption for a grain model consisting of two
separate grain populations -- carbonaceous grains and silicate grains.
With the grains approximated by homogeneous spheres with the
size distribution found by Weingartner \& Draine (2001;
hereafter WD01), this grain model is consistent with the observed
interstellar extinction law, 
the observed infrared emission from interstellar dust 
(Li \& Draine 2001, 2002),
and the
X-ray scattering halo observed around Nova Cygni 1992
(Draine \& Tan 2003).
The carbonaceous grains are assumed to be graphitic when the grains
are large, extending down to very small sizes with the smallest
grains being individual polycyclic aromatic hydrocarbon molecules.
The scattering is dominated by grains with radii $a\gtsim 100\Angstrom$,
containing $\gtsim 10^6$ atoms;
carbonaceous grains in this size range are modelled using the optical
properties of graphite.

The primary
objective of this paper is to calculate the X-ray scattering and absorption
properties of this dust model, and to make these results available for
comparison with observations.
The paper has three relatively independent parts, and readers may
wish to proceed directly to the material of interest to them.

The first part of this paper -- \S\ref{sec:dielec} --
is concerned with the dielectric functions
of graphite and MgFeSiO$_4$
from infrared to
X-ray energies. 
The resulting dielectric functions satisfy the Kramers-Kronig
relations as well as the oscillator strength sum rule.

The second part of the paper -- \S\S\ref{sec:XrayXsections}--\ref{sec:halos} 
-- examines the X-ray scattering properties of interstellar dust,
including structure near X-ray absorption edges.  
The angular distribution of the scattered X-rays is discussed:
the median scattering angle $\thetasm$ is found as
a function of energy, and a simple analytic approximation to the
differential scattering cross section is presented.
In \S\ref{sec:halos} we provide analytic approximations for the
intensity distribution in X-ray scattering halos for simple dust
density distributions.

The final part of the paper, \S\ref{sec:Xrayobs}, 
discusses observations of X-ray scattering halos, and values of
total scattering cross section derived from these observations.
We conclude that the observational situation is unclear at this time,
due to the combined effects of instrumental limitations and uncertainties
in the spatial distribution of the dust,
but the silicate/carbonaceous grain model appears to be consistent
with the overall body of observational data.
Future observations using extragalactic point sources could transcend
the uncertainties concerning the dust spatial distribution.

The principal results are summarized in \S\ref{sec:summary}

\section{Dielectric Function\label{sec:dielec}}

Our objective here is to obtain continuous complex dielectric functions
$\epsilon(\omega)=\epsilon_1+i\epsilon_2$
which can be used to calculate scattering and absorption by these
grains from the submm to hard X-rays.
For material at LTE (i.e., no population inversions), 
the dielectric function must have $\epsilon_2\geq0$.
Causality requires that 
the dielectric function must satisfy the Kramers-Kronig relations
(Landau \& Lifshitz 1960), in particular
\beq
\label{eq:KK}
\epsilon_1(\omega)=1+\frac{2}{\pi}P\int_0^\infty
\frac{\epsilon_2(x) x dx}{x^2-\omega^2}
~~~,
\eeq
where $P$ indicates that the principal value is to be taken.
We will proceed by first specifying $\epsilon_2(\omega)$ at all frequencies,
and then obtaining $\epsilon_1(\omega)$ using eq.\ (\ref{eq:KK}).

In addition to satisfying the Kramers-Kronig relations, the dielectric
function must obey the sum
rule (Altarelli et al.\ 1972)
\beq
\int_0^\infty \omega \epsilon_2(\omega) d\omega = 
\frac{2\pi^2e^2}{m_e} \sum_j n_j Z_j
~~~,
\eeq
where $n_j$ is the number density and $Z_j$ is 
the number of electrons in element $j$.
Thus we may define the effective number of electrons per molecule
\beq
\label{eq:neff}
n_{\rm eff}(\omega) \equiv \frac{m_e V_0}{2\pi^2e^2}\int_0^\omega x 
\epsilon_2(x)dx ~~~,
\eeq
where $V_0$ is the volume per molecule, and $n_{\rm eff}(\infty) =$ the
total number of electrons per molecule.

Ideally, $\epsilon_2(\omega)$ would be measured in the laboratory for the
materials of interest.
At energies below $\sim$20-30 eV the optical constants can be characterized
using transmission and ellipsometric studies, but 
calibrated experimental measurements are usually unavailable at
X-ray energies.
At high energies $\hbar\omega \gtsim 100 \eV$, 
$\epsilon_2(\omega)$ for a material
can be approximated by summing the atomic absorption cross sections
of the constituent atoms:
\beq
\label{eq:eps2_sigma}
\epsilon_2(\omega) \approx 
\frac{c}{\omega}\sum_j n_j \sum_s \sigma_{j,s}(\omega)
~~~,
\eeq
where $\sigma_{j,s}$ is the atomic absorption cross
section contributed by electronic shell $s$ of element $j$.
Eq. (\ref{eq:eps2_sigma}) assumes that $|\epsilon-1|\ll 1$, which is valid
at X-ray energies.

At energies well above the photoionization threshold for shell $s$,
atomic photoionization is to high momentum 
free electron states which will have
counterparts in the solid, and we can approximate
$\sigma_{j,s}$ by the atomic photoionization
cross section fitting functions 
$\sigma_{j,s}^{pi}(\omega)$ estimated for inner shell electrons 
by Verner \& Yakovlev (1995), and for outer-shell electrons by
Verner et al.\ (1996), as implemented in the Fortran routine
{\tt phfit2.f} (Verner 1996).
Near threshold, however, the photoabsorption cross section 
$\sigma_{j,s}(\omega)$ depends on
the band structure of the solid, leading to
``Near Edge X-ray Absorption Fine Structure'' (NEXAFS), which could
in principle permit 
identification
of interstellar grain materials through observations of
X-ray absorption and scattering
near absorption edges (Martin 1970; Woo 1995).

\subsection{Graphite}

We will assume that interstellar grains are constructed primarily
from two distinct substances: carbonaceous material and amorphous silicate
material.
In ultrasmall grains, containing less than $\sim10^5$ atoms, 
the carbonaceous material has the properties of
polycyclic aromatic hydrocarbon molecules.  Because of their small size,
scattering by these grains is negligible, and the IR-optical-UV absorption 
can be calculated using
absorption cross sections estimated for PAH molecules or ions
(Draine \& Li 2001; Li \& Draine 2001).

In the larger carbonaceous grains, the nature of the carbon material is
less certain.  The carbon atoms could be arranged in a graphitic
structure (pure $sp^2$ bonding) or there might be a mixture of
$sp^2$ (graphitic) and $sp^3$ (diamond-like) bonding, perhaps also with
some aliphatic (chainlike) hydrocarbon material as well.
We will assume that the optical response of the carbonaceous material
in the grains containing $\gtsim10^5$ atoms can be approximated by
graphite, with density $\rho=2.2\g\cm^{-3}$.

Graphite is anisotropic, with the crystal ``c axis'' 
normal to the basal plane.
In a cartesian coordinate system with $\hat{\bf z}\parallel c$, the
dielectric tensor is diagonal with eigenvalues
$(\epsilon_\perp,\epsilon_\perp,\epsilon_\parallel)$.

\begin{figure}[ht]
\centerline{
\epsfig{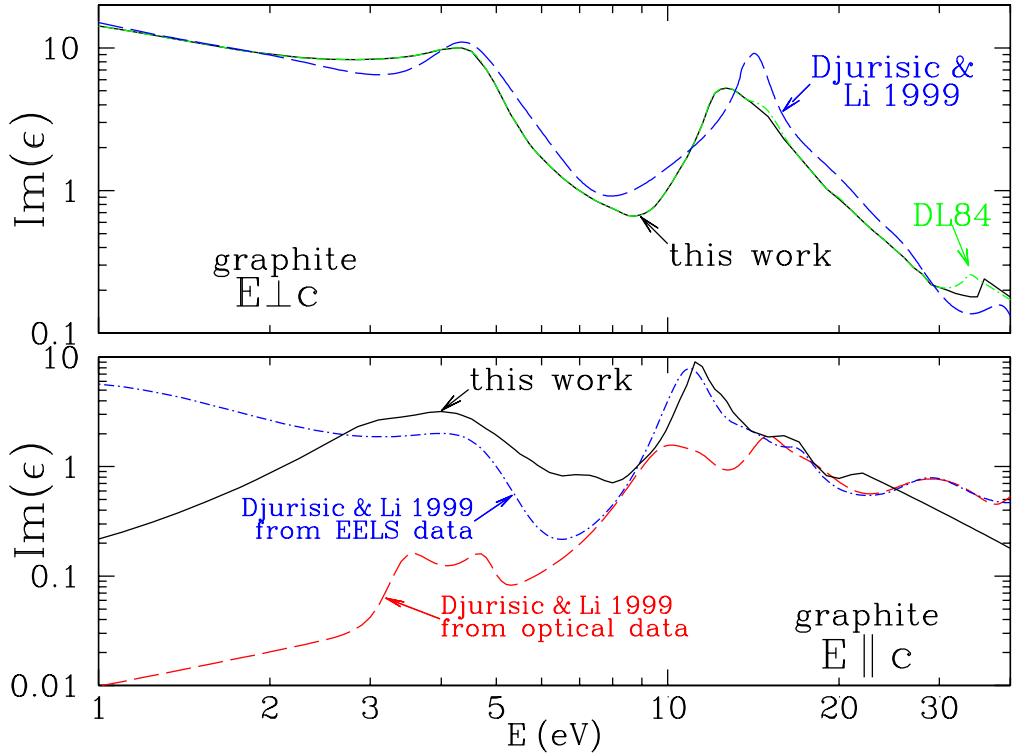}}
\caption{\footnotesize
	\label{fig:eps2_C_1_40}
	${\rm Im}(\epsilon)$ for graphite at $E<40\eV$ (see text).
	The $\epsilon_\perp$ adopted here is equal to $\epsilon_\perp$
	adopted by DL84 below 30 eV.
	Also shown, for comparison, are estimates for $\epsilon_\perp$ and
	$\epsilon_\parallel$ by Djurisic \& Li (1999).
	}
\end{figure}

The dielectric functions $\epsilon_\perp$ and $\epsilon_\parallel$ were
estimated by Draine \& Lee (1984, hereafter DL84) and
Laor \& Draine (1993).
For $E < 22 \eV$ we continue to use $\epsilon_{\parallel,2}$ from DL84,
while for $E > 22\eV$ we use eq. (\ref{eq:eps2_sigma}) to estimate
$\epsilon_{\parallel,2}$.
For $E < 35 \eV$ we use $\epsilon_{\perp,2}$ estimated by DL84,
while for $E > 35\eV$ we use eq. (\ref{eq:eps2_sigma}) to estimate
$\epsilon_{\perp,2}$.
Our final $\epsilon_{\perp,2}$ and $\epsilon_{\parallel,2}$ are shown
in Figures \ref{fig:eps2_C_1_40} and \ref{fig:eps_C}.
\begin{figure}[ht]
\centerline{
\epsfig{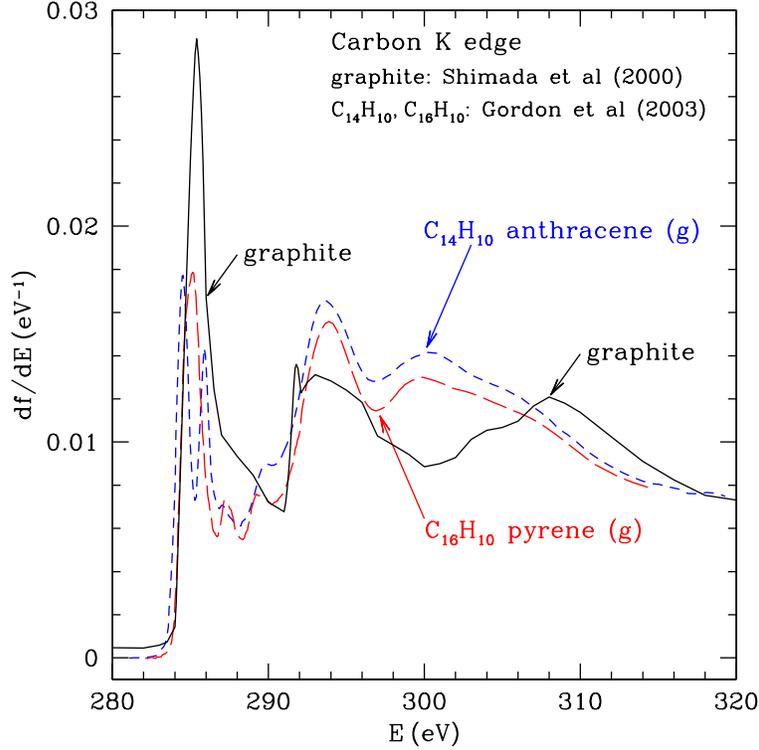}}
\caption{\footnotesize
	\label{fig:CKedge}
	Oscillator strength for absorption near the carbon K edge
	in gaseous anthracene and pyrene (Gordon et al.\ 2003), 
	and for graphite (Shimada et al.\ 2000, with normalization
	discussed in text).
	}
\end{figure}
\begin{figure}[ht]
\begin{center}
\epsfig{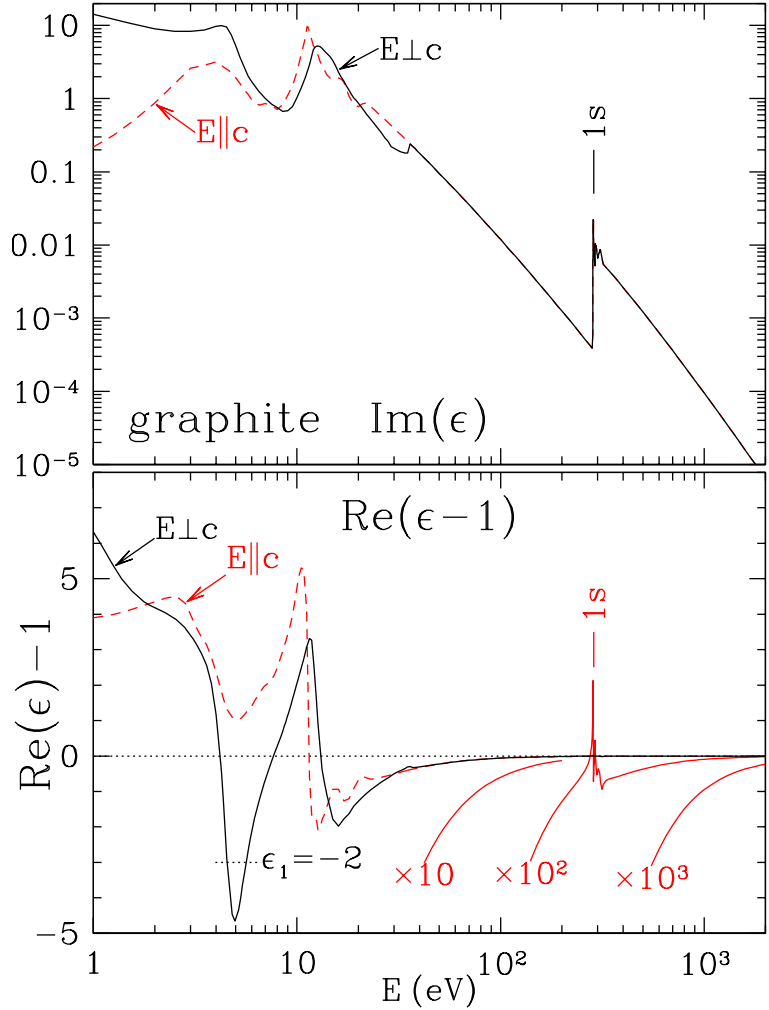}
\end{center}
\caption{\footnotesize
	\label{fig:eps_C}
	${\rm Im}(\epsilon)$ and ${\rm Re}(\epsilon-1)$
	for graphite from 1 eV to 2 keV.
	}
\end{figure}

Djurisic \& Li (1999) recently reestimated $\epsilon_\perp$ and
$\epsilon_\parallel$ for $\hbar\omega < 40\eV$ by fitting a model
to the available experimental evidence.
Their estimate for ${\rm Im}(\epsilon_\perp)$ is shown in 
Figure \ref{fig:eps2_C_1_40}.
We see that it is generally similar to the DL84 dielectric function,
although with a pronounced peak at $\sim14\eV$ in place of the
broader feature peaking at $\sim12\eV$ in the DL84 
estimate of $\epsilon_\perp$.

For $E\parallel c$, however, Djurisic \& Li concluded that optical 
ellipsometric
measurements are inconsistent with the dielectric function estimated
from electron energy loss spectroscopy (EELS).  
Djurisic \& Li obtained
$\epsilon_\parallel(\omega)$ from the optical measurements
and then independently from the EELS data.  
These two estimates 
for $\epsilon_\parallel$ -- which differ substantially -- are
shown in Figure \ref{fig:eps2_C_1_40}.
The Djurisic \& Li estimate based on EELS data is fairly similar to the
DL84 estimate for $\epsilon_\parallel$ (which was based in part on 
experimental papers using EELS).
Given the inconsistencies among the different experimental investigations,
$\epsilon_\parallel$ in the optical and UV should be regarded as
uncertain.

For carbon, the only X-ray feature is the carbon K edge.
We use the measured K edge X-ray absorption
profile for graphite from Shimada et al.\ (2000) for 282 - 310 eV.
Shimada et al.\ did not determine the absolute absorption strength.
We fix the amplitude such that the
oscillator strength between 282 and 320 eV is 0.38, and we smoothly
connect to join 
the Verner \& Yakovlev (1995) photoionization cross section at 320 eV.
In Figure \ref{fig:CKedge} we show the oscillator strength density
\beq
\frac{df}{dE} = \frac{m_e V_0}{\pi e^2h}\omega \epsilon_2(\omega)
\eeq
near the carbon K edge for graphite, and, for comparison, 
for gas phase anthracene and pyrene (Gordon et al.\ 2003).

In Figure \ref{fig:eps_C} we show $\epsilon_{\parallel,2}$ and
$\epsilon_{\perp,2}$ from the optical to the X-ray region.
Figure \ref{fig:eps_C} shows $\epsilon_{\parallel,1}$
and $\epsilon_{\perp,1}$ obtained using eq. (\ref{eq:KK}).

A useful check on the validity of the dielectric function is provided by
eq. (\ref{eq:neff}).
Figure \ref{fig:neff} shows that 
$n_{\rm eff}(\omega)\approx 4$ just below the onset of K shell absorption,
and $n_{\rm eff}(\omega\rightarrow\infty)\rightarrow 6$, as expected for
carbon.

\subsection{Silicate}

We will assume that the silicate material has an olivine composition,
Mg$_{2x}$Fe$_{2(1-x)}$SiO$_4$.
Mg and Fe are approximately equally abundant in the ISM, and both
reside primarily in interstellar grains.  It is therefore reasonable
to take the silicate grain composition to be MgFeSiO$_4$, although it is
possible that, for example, some of the Fe may be another chemical form.
MgFeSiO$_4$ olivine has a density 3.8$\g\cm^{-3}$, intermediate between the
densities of forsterite (Mg$_2$SiO$_4$, 3.27$\g\cm^{-3}$)
and fayalite (Fe$_2$SiO$_4$, 4.39$\g\cm^{-3}$),
with a molecular volume $V_0=7.5\times10^{-23}\cm^3$.

For $h\nu < 18\eV$ we will adopt $\epsilon_2$ 
previously obtained by Draine \& Lee (1984) for ``astronomical
silicate'', but
with the following modifications to $\epsilon_2$:
\begin{enumerate}
\item the crystalline olivine feature at $\lambda^{-1}=6.5\micron^{-1}$ -- not
seen in interstellar extinction or polarization (Kim \& Martin 1995) -- has
been excised with the oscillator strength redistributed over frequencies
between 8 and 10$\micron^{-1}$ (Weingartner \& Draine 2001)
\item at $\lambda > 250\micron$, $\epsilon_2(\omega)$ has been modified
slightly, as described by Li \& Draine (2001).
For $250 < \lambda < 1100\micron$ the revised $\epsilon_2$ is within
$\pm12\%$ of $\epsilon_2$ adopted by DL84.
\end{enumerate}

Above 30 eV, we use eq. (\ref{eq:eps2_sigma}) to estimate $\epsilon_2$
from atomic photoabsorption cross sections, except near absorption edges
(see below).
Between 18 and 30 eV, $\epsilon_2$ is chosen to provide a smooth join
between $\epsilon_2$ from DL84 and $\epsilon_2$ estimated from the
atomic photoabsorption cross sections.

The threshold energy for photoionization from the K shell of atomic oxygen is
544.0~eV for ionization to O~II~$1s2s^22p^4(^4{\rm P})$,
and 548.9~eV for ionization to O~II~$1s2s^22p^4(^2{\rm P})$;
the strong $1s-2p$ absorption line,
with FWHM $\approx 0.14 \eV$, lies at 527.0~eV (Stolte et al.\ 1997).
From the theoretical line profile of McLaughlin \& Kirby (1998), the
$1s-2p$ transition has an oscillator strength $f\approx 0.10$.

X-ray spectroscopy of several galactic X-ray sources 
using the {\it Chandra X-Ray Observatory}
has detected a strong and narrow absorption line at 527.5~eV 
which must be the O~I $1s-2p$
transition, and a nearby absorption feature at 530.8~eV,
with FWHM $\sim$1.0eV 
(Paerels et al.\ 2001; Schulz et al.\ 2002; Takei et al.\ 2002).
Paerels et al.\ and Schulz et al.\ suggest that the 530.8 eV feature is due to
iron oxides, possibly Fe$_2$O$_3$.
However, O~II is expected to have $1s-2p$ absorption at approximately this
energy, and the interstellar O~II abundance is probably large enough
that the resulting absorption feature should be conspicuous.
It therefore seems likely that the observed narrow feature at 530.8~eV
is (at least primarily) O~II~$1s-2p$.

In the absence of published X-ray absorption spectra for amorphous silicates,
we estimate the oxygen K edge absorption using measurements on
crystalline olivines.
Li et al.\ (1995) have measured Mg K edge and Si K edge
absorption for
forsterite Mg$_2$SiO$_4$, and Henderson et al.\ (1995) have
measured Fe K edge absorption in fayalite Fe$_2$SiO$_4$.
We adopt these profiles for the corresponding K edges in
amorphous olivine MgFeSiO$_4$, with the
absorption profile strengths adjusted to match the 
Verner et al.\ photoionization 
cross sections for the corresponding atoms well above threshold.

Fe L$_{2,3}$ edge spectra for a number of minerals 
have been studied recently by van Aken \& 
Liebscher (2002).
The L$_2$ edge corresponds to
the remaining $2p^5$ electrons in a $^2$P$_{3/2}$ term,
and the L$_3$ edge when they are in a $^2$P$_{1/2}$ term.
The Fe in olivine is 100\% Fe$^{2+}$.
The spin-orbit splitting for Fe$^{2+}$ produces a 12.8 eV separation between
the L$_2$ and L$_3$ maxima (van Aken \& Liebscher 2002), with
the L$_2$ peak at 721.3 eV and the L$_3$ peak at 708.1 eV.
Following van Aken \& Liebscher, we model the near-edge absorption in
olivine by
\begin{eqnarray}
\sigma_{L_{2,3}} &=& \sigma_0 \left\{ 
	\frac{2}{3\pi}\left[
			\arctan\left(\frac{\pi}{w_a}(E-E_a)\right)+
			\frac{\pi}{2}
			\right]
	+\frac{1}{3\pi}\left[
			\arctan\left(\frac{\pi}{w_b}(E-E_b)\right)+
			\frac{\pi}{2}
			\right]
	\right.
	\nonumber
	\\
	&&
	\left.
	+ \sum_{i=1}^4 A_i\exp\left[-\left(\frac{E-E_j}{w_j}\right)^2\right]
	\right\}~~~,
	\label{eq:L23}
\end{eqnarray}
with $E_a=708.65\eV$, $E_b=721.65\eV$, 
and $w_a=w_b=1\eV$.
The gaussian components are located at
$E_1=707.8\eV$, $E_2=710.5\eV$, $E_3=720.6\eV$, and $E_4=723.3\eV$.
van Aken \& Liebscher do not give values for the widths of the gaussians, but
inspection of their Fig.\ 1 suggests 
$w_1=w_2=w_3=w_4=1.25\eV$.
Minerals with 100\% Fe$^{2+}$ have $A_1/A_2\approx A_3/A_4\approx 5$,
and $A_3/A_1\approx 0.28$.
For 100\% ferrous Fe we estimate $A_1\approx 6$ from the spectrum
of ilmenite FeTiO$_3$ in van Aken, Liebscher \& Styrsa (1998).
To determine the normalization $\sigma_0$, we require that eq.\ (\ref{eq:L23})
match the Verner et al.\ cross section at 800~eV.

For some absorption edges no measurements are available for appropriate
minerals.
Since the near-edge absorption is proportional to the density of electronic
states above the Fermi surface,
the near edge spectra for
different elements in a compound show considerable similarity in the
dependence on energy, but with displacements in energy due to the
different binding energies of the electron being excited.
We use the Si K edge in forsterite 
Mg$_2$SiO$_4$
(Li et al.\ 1995), shifted in energy by $\Delta E=-1308\eV$
[the difference in K edge ionization thresholds for 
Si~I (1846 eV)  and O~I (538 eV)] to determine 
the O K edge absorption profile up to 575~eV.
This results in onset of O 1s absorption at $\sim$527.8 eV, and
an O 1s absorption peak at 537.6 eV.
The absolute absorption coefficient is fixed by requiring that the
absorption at 575~eV match that calculated from the atomic photoionization
cross sections.\footnote{%
	Absorption near the O K edge in SiO$_2$ has been measured 
	by Marcelli et al.\ (1985), who find an absorption peak
	at 540~eV in both $\alpha-$quartz and glassy SiO$_2$.
	EELS studies by Wu et al.\ (1996) and Sharp et al.\ (1996)
	show the peak at 535 eV for quartz, while
	Garvie et al.\ (2000) use EELS to locate the quartz peak 
	at 537.5~eV, with the 
	absorption edge located at $\sim 535\eV$.
	The SiO$_2$
	polymorphs $\alpha$-quartz, coesite, and stishovite have their
	absorption peaks within $\sim 1\eV$ of one another (Wu et al.\ 1996),
	and likewise their absorption edges agree to within $\sim 1\eV$.%
	}

A similar approach is taken for the other absorption edges,
using measured profiles for Mg K edge and Si K edge absorption in
forsterite Mg$_2$SiO$_4$ (Li et al.\ 1995),
and Fe K edge absorption in fayalite Fe$_2$SiO$_4$ (Henderson et al.\ 1995)
(see Table \ref{tab:edge_profs}).

Although this procedure is not expected to provide
accurate estimates of the near-edge absorption spectra for amorphous
silicates,
the provisional near-edge absorption profiles so obtained will
provide a realistic example of the kind of near-edge absorption and scattering
expected
from interstellar grains.

\begin{table}[h]
\begin{center}
{\footnotesize
\caption{X-ray Edge Absorption Parameters\label{tab:edge_profs}}
\begin{tabular}{c c c c c c c c}
\tableline\tableline
material&shell
	& $E_{min}$\tablenotemark{a}
	& $E_{peak}$\tablenotemark{b}
	& $\sigma_{peak}$\tablenotemark{c}
	& adopted
	& $\Delta E$\tablenotemark{d}&\\
	&&(eV)	&(eV)	&Mb	& profile	&(eV)	&ref\\
\tableline
graphite&C~$1s~(K)$	&282	&285.4	&3.84
	& graphite
	&0
	&\tablenotemark{e}\\
olivine&O$1s~(K)$	&527.8	&537.6	&1.78
	&Mg$_2$SiO$_4$ Mg K
	&-1308.0	
	&\tablenotemark{f}\\
olivine&Mg$1s~(K)$	&1300.8 &1310.6 &0.80
	&Mg$_2$SiO$_4$ Mg K
	&0	
	&\tablenotemark{f}\\
olivine&Mg~$2s~(L_1)$	&83.8	&93.6	&2.07
	&Mg$_2$SiO$_4$ Mg K
	&-1752.0
	&\tablenotemark{f}\\
olivine&Mg~$2p~(L_{2,3})$	&44.7	&54.5	&15.7
	&Mg$_2$SiO$_4$ Mg K
	&-1791.1	
	&\tablenotemark{f}\\
olivine&Si~$1s~(K)$	&1835.8	&1845.6	&0.50
	&Mg$_2$SiO$_4$ Si K
	&0
	&\tablenotemark{f}\\
olivine&Si~$2s~(L_1)$	&145.8	&155.6	&1.67
	&Mg$_2$SiO$_4$ Si K
	&-1690.0
	&\tablenotemark{f}\\
olivine&Si~$2p~(L_{2,3})$	&95.8	&105.6	&18.0
	&Mg$_2$SiO$_4$ Si K
	&-1740.0	
	&\tablenotemark{f}\\
olivine&Fe~$1s~(K)$	&7105	&7123	&.0544
	&Fe$_2$SiO$_4$ Fe K
	&0	
	&\tablenotemark{g}\\
olivine&Fe~$2s~(L_1)$	&838	&856	&0.186
	&Fe$_2$SiO$_4$ Fe K
	&-6267.	
	&\tablenotemark{g}\\
olivine&Fe~$2p(L_2)$	&705	&720.6	&2.46
	&Fe$^{2+}$ minerals Fe L$_2$
	&0	
	&\tablenotemark{h}\\
olivine&Fe~$2p(L_3)$	&705	&707.8	&6.30
	&Fe$^{2+}$ minerals Fe L$_3$
	&0	
	&\tablenotemark{h}\\
olivine&Fe~$3s~(M_1)$	&85	&103	&0.338
	&Fe$_2$SiO$_4$ Fe K
	&-7020	
	&\tablenotemark{g}\\
olivine&Fe~$3p~(M_{2,3})$	&47	&65	&1.41
	&Fe$_2$SiO$_4$ Fe K
	&-7058	
	&\tablenotemark{g}\\
\tableline
\end{tabular}
\tablenotetext{a}{Energy at onset of absorption}
\tablenotetext{b}{Energy at peak absorption}
\tablenotetext{c}{Peak absorption cross section/atom 
	contributed by this shell.}
\tablenotetext{d}{Energy shift relative to adopted profile.}
\tablenotetext{e}{Shimada et al.\ (2000)}
\tablenotetext{f}{Li et al.\ (1995)}
\tablenotetext{g}{Henderson et al.\ (1995)}
\tablenotetext{h}{van Aken \& Liebscher (2002)}
}
\end{center}
\end{table}

\begin{figure}[ht]
\begin{center}
\epsfig{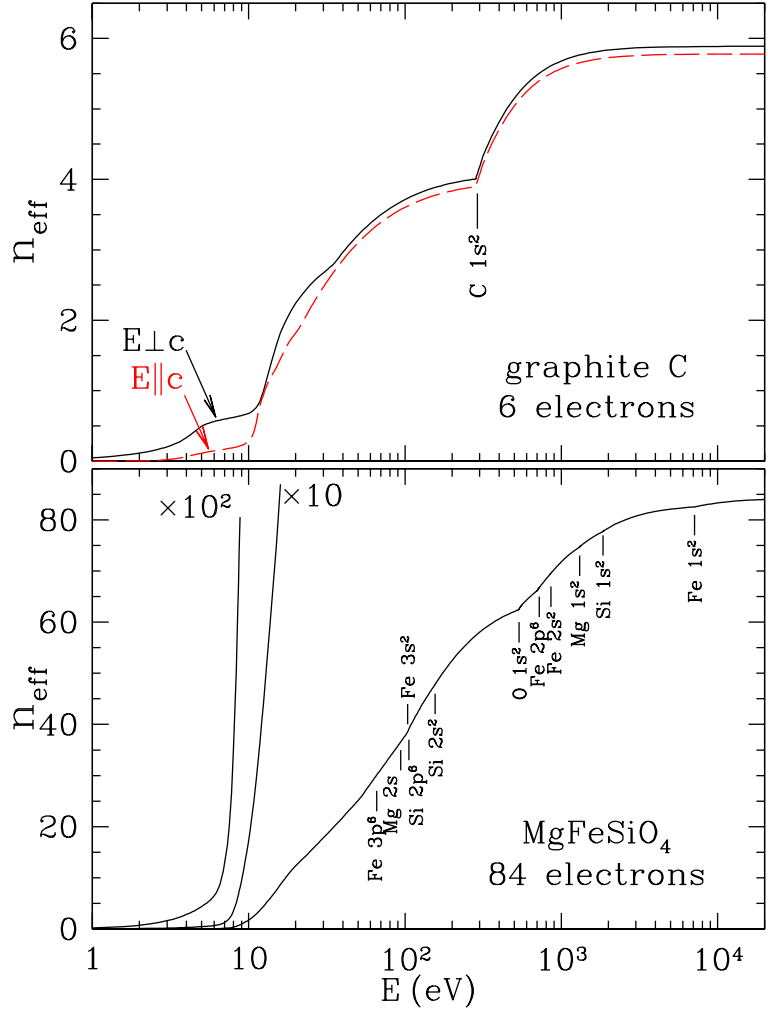}
\end{center}
\caption{\footnotesize
	\label{fig:neff}
	Effective number of electrons $n_{\rm eff}$ as a function
	of photon energy for graphite and MgFeSiO$_4$.
	}
\end{figure}

\begin{figure}[ht]
\begin{center}
\epsfig{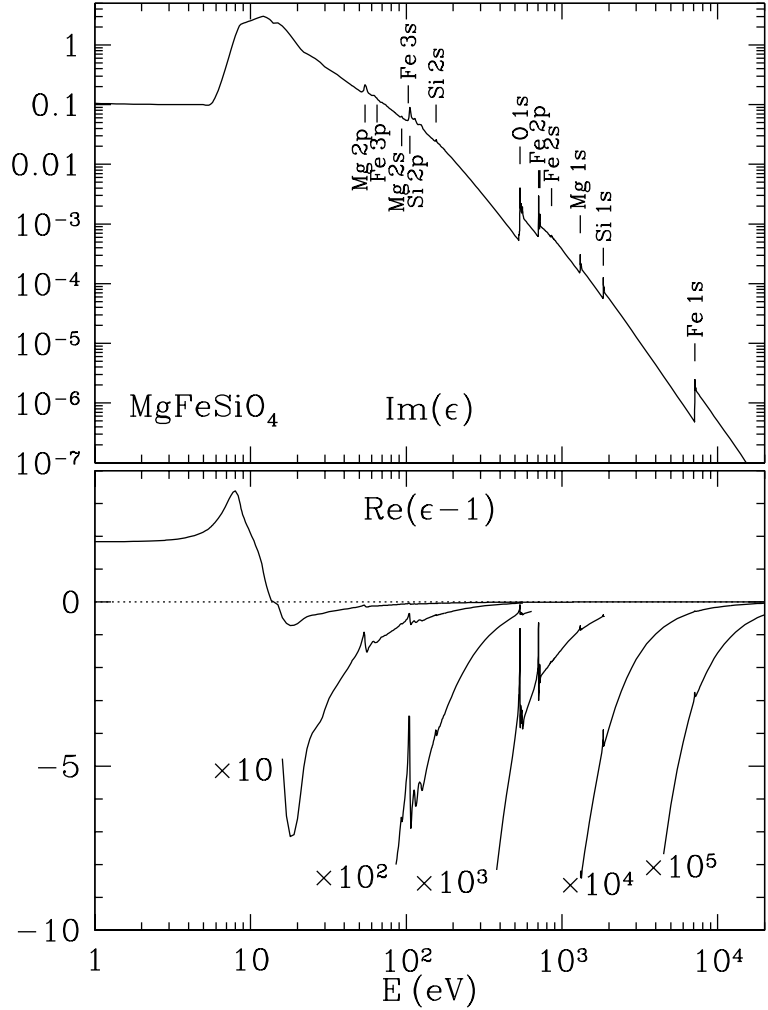}
\end{center}
\caption{\footnotesize
	\label{fig:eps_sil}
	${\rm Im}(\epsilon)$ and ${\rm Re}(\epsilon-1)$ for
	silicate MgFeSiO$_4$.
	}
\end{figure}

Figure \ref{fig:eps_sil} shows ${\rm Re}(\epsilon)$ obtained from 
${\rm Im}(\epsilon)$
using eq.\ (\ref{eq:KK}).
Figure \ref{fig:neff} provides a check on the adopted dielectric function:
$\lim_{\omega\rightarrow\infty}n_{\rm eff}=84.01$, in agreement
with the 
value of 84 expected theoretically for
MgFeSiO$_4$.


\section{Cross Sections for X-Ray Absorption and Scattering
	\label{sec:XrayXsections}}

Adopting the dielectric functions discussed in
\S\ref{sec:dielec}, we calculate the scattering and absorption for
the WD01 dust grain mixture.
We use the Mie scattering theory program of Wiscombe (1980, 1996) for
$x=2\pi a/\lambda < 2\times10^4$, and anomalous diffraction theory
(van de Hulst 1957) for $x > 2\times10^4$.
Anomalous diffraction theory provides an excellent approximation at
the X-ray energies where $x=5067(a/\micron)(E/\keV) > 2\times10^4$
(see Figure 7 of Draine \& Tan 2003).

\begin{figure}[h]
\centerline{\epsfig{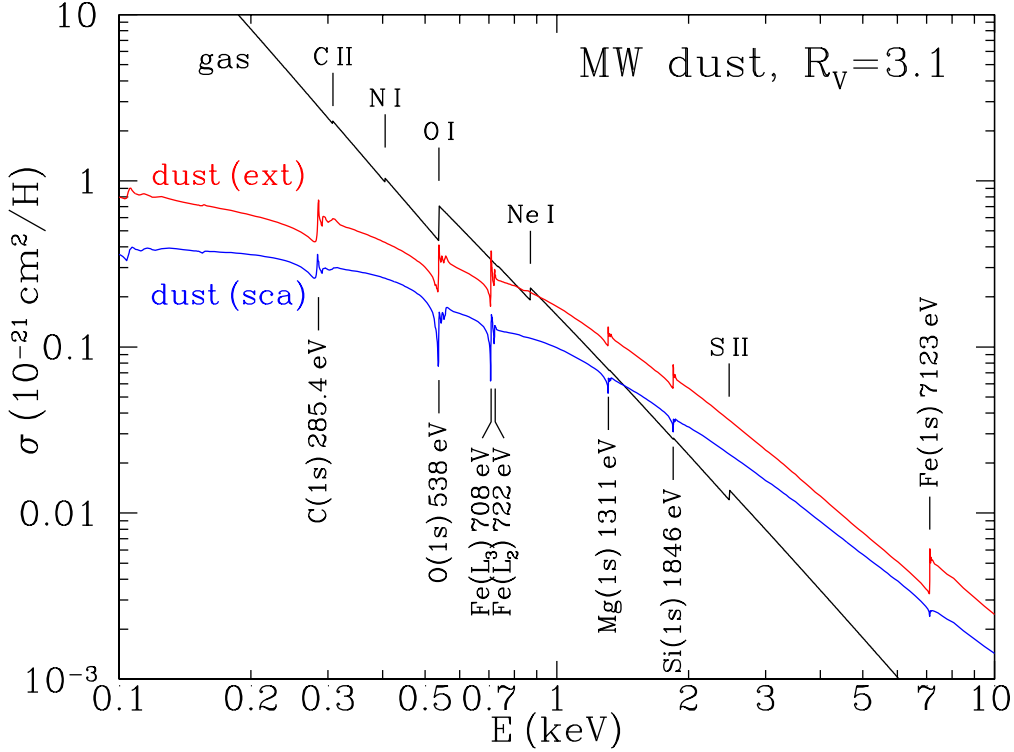}}
\caption{\footnotesize
	\label{fig:Xrayextsca}
	X-ray extinction and scattering cross section per H nucleon
	due to dust, and absorption due to gas.
	Data available at http://www.astro.princeton.edu/$\sim$draine/dust.html
	}
\end{figure}

\begin{figure}[h]
\centerline{\epsfig{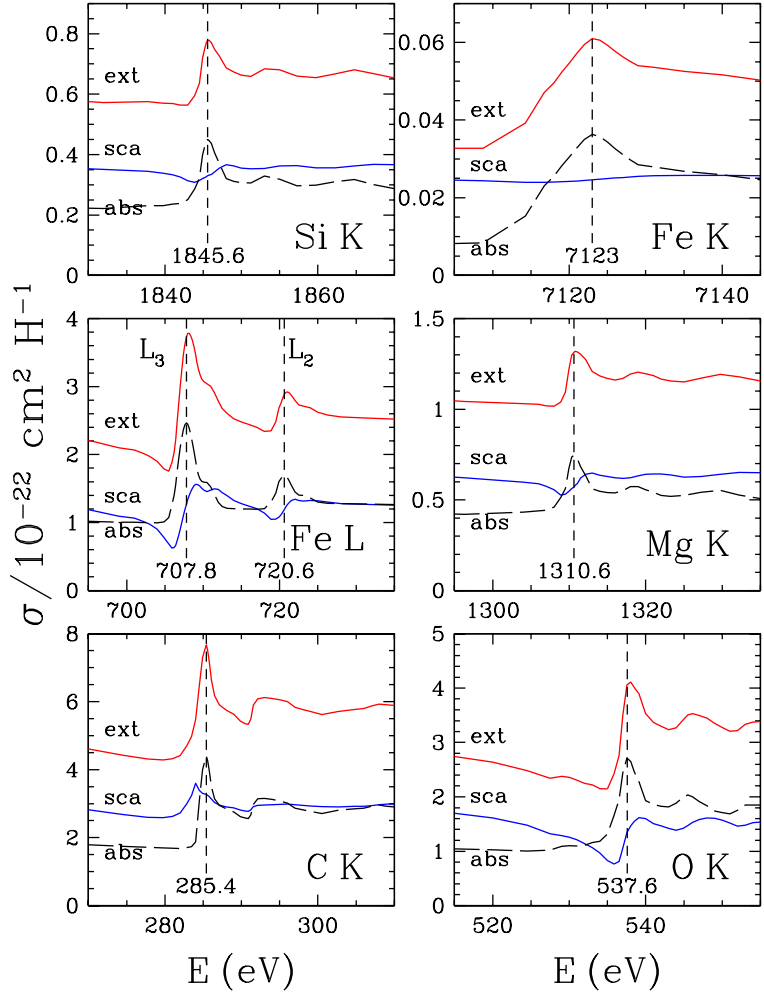}}
\caption{\footnotesize
	\label{fig:Xrayextsca_edge}
	X-ray extinction and scattering cross section per H nucleon
	near major absorption edges.
	}
\end{figure}

The scattering and extinction cross section calculated for the dust mixture
is shown in
Fig.\ \ref{fig:Xrayextsca}
for 0.1--10~keV X-rays, with the six strongest absorption edges
shown in Fig.\ \ref{fig:Xrayextsca_edge}.
Also shown is the absorption per H atom due to gas-phase absorption
calculated using {\tt phfit2.f} (Verner 1996), with interstellar
gas-phase abundances.
At energies $13.6 < E \ltsim 250\eV$, 
absorption by neutral H and He is very strong,
making it difficult to observe the dust absorption
and scattering.  Above 250 eV, however, observations
of extinction and scattering by dust become feasible for suitably bright
sources on sightlines with sufficient dust columns.
At $E\gtsim 800\eV$ the extinction is primarily due to dust grains.

As seen in Figure \ref{fig:Xrayextsca_edge}, 
the calculated scattering cross sections
show conspicuous structure in the vicinity
of the major absorption edges.
This occurs because at X-ray energies ${\rm Re}(\epsilon-1)$ tends
to be negative, and an absorption feature increases
${\rm Re}(\epsilon)$
(reducing $|\epsilon-1|$)
just below the absorption feature, 
and decreases ${\rm Re}(\epsilon)$ (increasing $|\epsilon-1|$) just above
the feature.
Since the scattering is approximately proportional to
$|\epsilon-1|^2$, this results in a reduction in scattering below an
absorption feature, and an increase in scattering above.\footnote{
	Takei et al.\ (2002) estimate that the dust scattering
	cross section would be reduced at energies just above
	the O~K edge.
	We find, to the contrary, that the dust scattering
	cross section is {\it increased} just above the O~K edge --
	see Fig.\ \ref{fig:Xrayextsca_edge}.
	}
This argument
applies to 5 of the 6 absorption edges in Fig.\ \ref{fig:Xrayextsca_edge};
the exception is the C K edge, for which $\epsilon -1$ actually becomes
positive below the absorption edge, with a local peak in scattering just
below the K edge.


For the lower-energy absorption edges (C~K, O~K, Fe~L$_{2,3}$)
there is significant variation in the scattering
cross section near the absorption edge.  As a result, the
extinction profile is not the same as the absorption profile. 
In the case of the O K edge, the extinction peak is at 538.0 eV,
whereas the absorption peak is at 537.6 eV.
The energy-dependence of the scattering optical depth can be determined
by dividing the spectrum of the scattered X-rays by the
spectrum of the point source component.  If this can be done with
a signal-to-noise ratio $\gtsim10$ for $\sim2\eV$ bins, one should observe
the structure in $\sigma_{\rm sca}$ seen near
285, 538, and 708~eV in Figure \ref{fig:Xrayextsca_edge}.

\begin{figure}[h]
\centerline{\epsfig{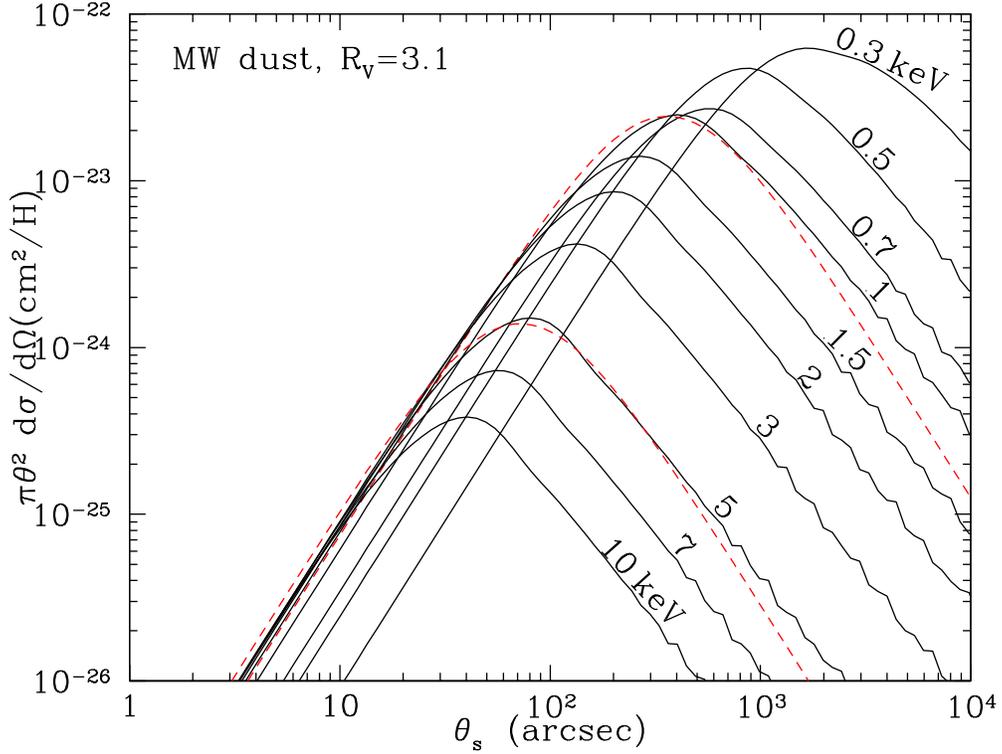}}
\caption{\footnotesize
	\label{fig:Xraysca}
	$2\pi\theta^2d\sigma/d\Omega$ versus scattering angle $\theta_s$
	at selected energies.  This function peaks at
	approximately the median scattering angle.
	The broken curves show eq.\ (\ref{eq:sigma<theta}) for
	$E=1.0$ and $5.0\keV$.
	}
\end{figure}

At X-ray energies, the dielectric functions of grain materials
become close to unity (see Figures \ref{fig:eps_C} and \ref{fig:eps_sil}),
the wavelength is small compared to the typical grain radius,
and the grains are very strongly forward-scattering.
The quantity
\beq
\frac{d\sigma_{\rm sca}}{d\ln\Omega} = 
\Omega \frac{d\sigma}{d\Omega} = 2\pi (1-\cos\theta) \frac{d\sigma}{d\Omega}
\approx \pi \theta^2 \frac{d\sigma}{d\Omega} ~~~
\eeq
is proportional to the number of scattered photons per logarithmic
interval of scattering angle; the location of the peak shows
the ``typical'' scattering angle.
Figure \ref{fig:Xraysca} shows $\pi\theta^2d\sigma/d\Omega$ for the
WD01 dust mixture for selected energies from 0.3 keV to 10 keV.

\begin{figure}[h]
\centerline{\epsfig{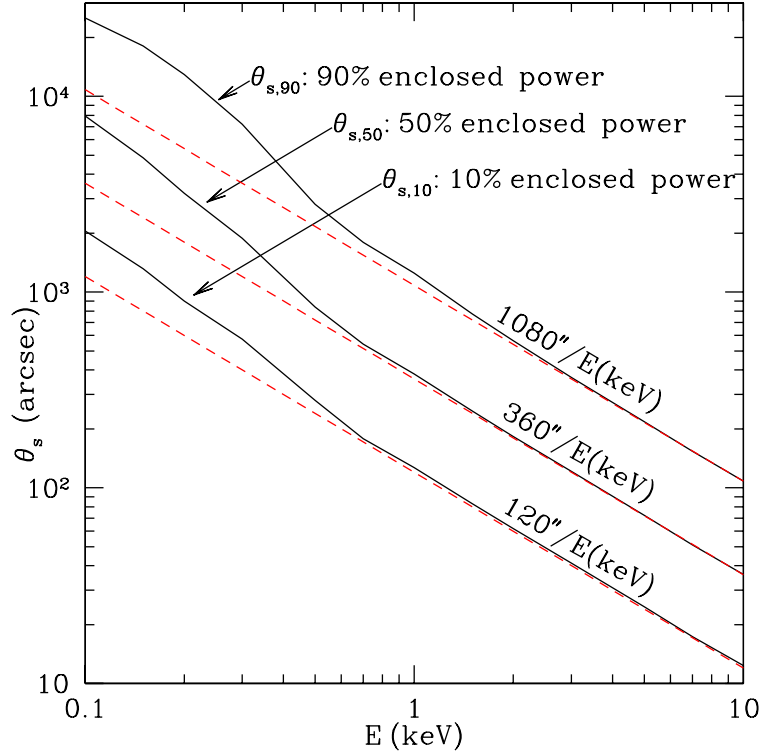}}
\caption{\footnotesize
	\label{fig:thetam}
	Median scattering angle $\thetasm$ as a function of energy
	for the WD01 grain model.
	Also shown are scattering angles $\thetasten$ and
	$\thetasninety$ for 10\% and 90\% enclosed
	power.
	Broken lines show asymptotic behavior for $E \gtsim 1 \keV$.
	}
\end{figure}

Let $\thetasm(E)$ be the median scattering angle
for photons of energy $E$.
Figure \ref{fig:thetam} shows $\thetasm(E)$ for
the WD01 dust mixture; comparison with Fig.\ \ref{fig:Xraysca} shows that,
as expected,
the median scattering angle
$\thetasm$ is very nearly the same as the angle where
$\pi\theta^2d\sigma/d\Omega$ peaks.
Also shown in Fig.\ \ref{fig:thetam}
are the 10th and 90th percentile scattering angles, 
$\thetasten$ and $\thetasninety$,
corresponding to 10\% and 90\% enclosed
power.
Note that to a very good approximation,
\beq
\thetasten\approx \frac{\thetasm}{3} ~~~,~~~\thetasninety\approx3\thetasm
~~~.
\eeq
For $E\gtsim 0.5\keV$, 
the median scattering angle for the WD01 dust mixture
can be approximated by
\beq
\label{eq:thetasm}
\thetasm \approx 360\arcsec \left(\frac{\keV}{E}\right) ~~~.
\eeq
The median scattering angle for a circular aperture of diameter $d$ is
$0.53\lambda/d$ (Born \& Wolf 1999), 
so equation (\ref{eq:thetasm}) corresponds to
the median scattering angle for an aperture of radius $0.19\micron$,
consistent with the size of the
grains which dominate the
visual extinction and polarization of starlight, account for
most of the interstellar grain mass, and are expected to dominate
the X-ray scattering.
The smaller grains, while more numerous, make only a minor contribution
to scattering at $E\gtsim 0.5\keV$.  Note that at lower energies, the
median scattering angle rises above the approximation (\ref{eq:thetasm}),
due in part to the increasing importance of smaller grains (which
contribute most of the geometric cross section of the grain population)
at these energies.

For this dust model, the differential scattering cross section
can be approximated by
the simple analytic form
\beq
\label{eq:dsigdom_approx}
\frac{d\sigma}{d\Omega} \approx 
\frac{\sigma_{\rm sca}}{\pi\thetasm^2}
\frac{1}{\left[1+(\theta/\thetasm)^2\right]^2}
~~~,
\eeq
with the total cross section for scattering angles $<\theta$
\beq
\label{eq:sigma<theta}
\sigma_{\rm sca}(<\theta) = \sigma_{\rm sca} 
\frac{(\theta/\thetasm)^2}{1+(\theta/\thetasm)^2}
~~~.
\eeq
Eq.\ (\ref{eq:sigma<theta}) reproduces the empirical
result that $\thetasten=\thetasm/3$, and $\thetasninety=3\thetasm$.
The approximation (\ref{eq:dsigdom_approx}) is plotted in Fig.\
\ref{fig:Xraysca} for $E=1.0$ and 5.0~\keV, showing that it does
indeed provide a good fit.

\section{X-Ray Scattering Halos: Models
	\label{sec:halos}}

\subsection{Models}
For a point source at distance $D$, scattering by dust on the sightline
a distance
$r=xD$ from the observer produces a scattered halo around the
point source (see, e.g., Draine \& Tan 2003) 
with the halo angle $\thetah$ related to the
scattering angle $\thetas$ through
\beq
\label{eq:thetah_thetas}
\thetah \approx (1-x)\thetas
~~~.
\eeq
Let $\Nhalo$ be the total flux of singly-scattered photons, 
and $\Nhalo(<\thetah)$
be the flux of photons at halo angles $<\thetah$.
Define the fraction of halo photons interior to $\thetah$:
\beq
g(\theta_h) \equiv \frac{\Nhalo(<\thetah)}{\Nhalo} ~~~.
\eeq
If the dust density is assumed to be plane-parallel perpendicular to
the sightline, then for the small-angle scattering appropriate
to X-ray energies, the scattering halo is given by
\beq
g(\thetah)\equiv
\int_0^1 dx ~\tilde{\rho}(x) 
\frac{\sigma_{\rm sca}(<\theta_h/(1-x))}{\sigma_{\rm sca}}
~~~,
\label{eq:gint}
\eeq
where the dimensionless dust density
\beq
\tilde{\rho}(x) \equiv \frac{\rho(xD)}{\int_0^1\rho(xD)dx}
~~~,
\eeq
where $\rho(r)$ is the dust density along the sightline at distance
$r$ from the observer.

Because the differential scattering cross section for the WD01 dust
mixture can be approximated by 
eq.\ (\ref{eq:dsigdom_approx},\ref{eq:sigma<theta}), we have
\beq
g(\thetah)\approx
\int_0^1\tilde{\rho}(x) \left[1+(1-x)^2(\thetasm/\thetah)^2\right]^{-1}
~~~.
\eeq

\begin{figure}[h]
\centerline{\epsfig{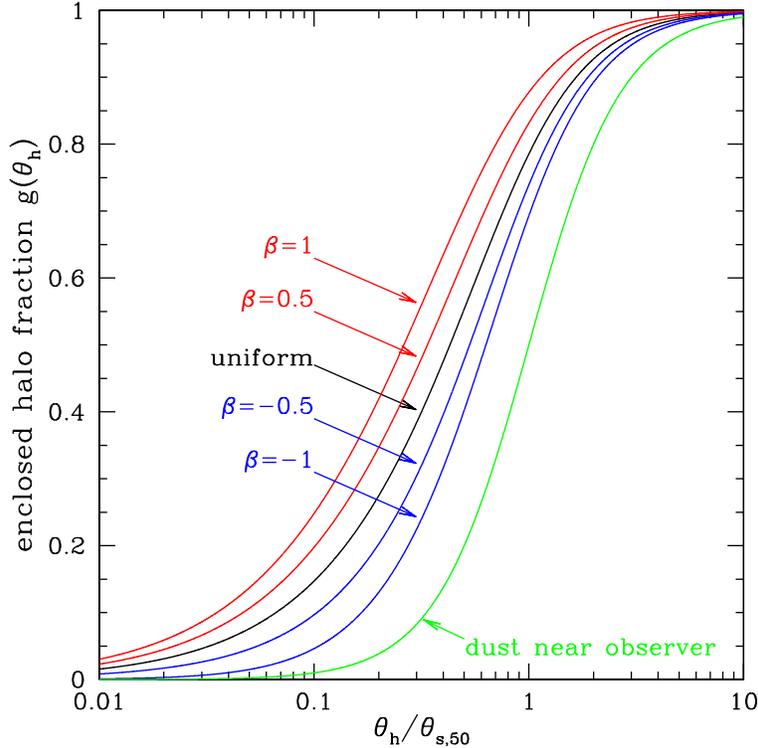}}
\caption{\footnotesize
	\label{fig:gfunc}
	Fraction of the single-scattering halo falling within
	halo angle $\theta_h$, as a function of $\theta_h/\theta_{s,50}$,
	where $\theta_{s,50}$ is the median scattering angle.
	}
\end{figure}

If the scattering is by a single sheet of dust at distance $x_dD$, then
$\tilde{\rho}(x)=\delta(x-x_d)$ and
\beq
\label{eq:g_sheet}
g(\theta_h) \approx \frac{(\thetah/\thetahm)^2}{1+(\thetah/\thetahm)^2}
~~~,~~~
\thetahm = (1-x_d)\thetasm
~~~.
\eeq
For a uniform dust density gradient (with $\beta=0$ corresponding to
uniformly-distributed dust)
\beq
\tilde{\rho}(x) = (1-\beta) + 2\beta x  ~~~~~(-1 \leq \beta \leq 1)
~~~,
\eeq
eq.\ (\ref{eq:gint}) can be integrated to obtain
\beq
\label{eq:g(theta,beta)}
g(\theta_h) = (1+\beta)\frac{\thetah}{\thetasm}\arctan(\thetasm/\thetah)
-\beta \left(\frac{\thetah}{\thetasm}\right)^2
\ln\left[1+(\thetasm/\thetah)^2\right]
~~~;
\eeq
$g(\theta_h)$ is plotted in Fig.\ \ref{fig:gfunc} for 5 cases:
$\beta=-1$ (1/4 of the dust between $x=0.5$ and 1);
$\beta=-.5$ (3/8 of the dust between $x=0.5$ and 1);
$\beta=0$ (uniform dust);
$\beta=0.5$ (5/8 of the dust between $x=0.5$ and 1);
$\beta=1$ (3/4 of the dust between $x=0.5$ and 1).
Also plotted is the case where the dust is all at $x\ll 1$,
with $\thetah=\thetas$.
For the above dust distributions, 
Table \ref{tab:thetah_values} gives the halo angles $\theta_{h,10}$,
$\theta_{h,50}$, $\theta_{h,90}$ enclosing 10\%, 50\%, and 90\% of the
halo power for single-scattering.

\begin{table}[h]
\caption{\label{tab:thetah_values}
	Halo Structure Parameters for WD01 Dust}
{\footnotesize
\begin{tabular}{cccccc}
\tableline\tableline
&\multicolumn{5}{c}{dust density distribution}\\
	&$\beta=-1$	
		&$\beta=-0.5$
			&uniform	
				&$\beta=0.5$	
					&$\beta=1$\\
\tableline
$\thetahten/\thetasm$
	&0.1663	&0.1032	&.0664	&.0467	&.0353\\
$\thetahm/\thetasm$
	&0.631	&0.530	&0.429	&0.337	&0.262\\
$\thetahninety/\thetasm$
	&2.084	&1.882	&1.660	&1.413	&1.139\\
$\thetahten\times(E/\keV)$
	&$59.9\arcsec$
		&$37.2\arcsec$
			&$23.9\arcsec$
				&$16.8\arcsec$
					&$12.7\arcsec$\\
$\thetahm\times(E/\keV)$
	&$227\arcsec$
		&$191\arcsec$
			&$154\arcsec$
				&$121\arcsec$
					&$94.3\arcsec$\\
$\thetahninety\times(E/\keV)$
	&$750\arcsec$
		&$678\arcsec$
			&$598\arcsec$
				&$509\arcsec$
					&$410\arcsec$\\
\tableline
\end{tabular}
}
\end{table}

\section{X-Ray Scattering Halos: Observations\label{sec:Xrayobs}}

The total cross section for X-ray scattering can be measured by imaging the
scattered X-ray halo.
The flux of scattered photons, $\Nhalo$,
is related to the flux in the point-source component, $\Nptsrc$,
by $\Nptsrc=(\Nptsrc+\Nhalo)\exp(-\tausca)$, so
\beq
\label{eq:tausca}
\tausca=\ln(1+\Nhalo/\Nptsrc) = -\ln\left(1-\fhalo\right) ~~~,
\eeq
\beq\fhalo\equiv\frac{\Nhalo}{\Nhalo+\Nptsrc} ~~~;
\eeq
eq.\ (\ref{eq:tausca}) is valid even when
multiple scattering takes place.
Absorption (by dust or gas) has a negligible effect on
$\Nhalo/\Nptsrc$,
because unscattered and scattered photons 
are affected essentially equally.
Estimation of $\tausca$ from (\ref{eq:tausca}) requires determination of
the flux of scattered photons $\Nhalo$ integrated over all halo angles.
If the halo flux is measured only for halo angles 
$\theta_1 < \thetah < \theta_2$, the total halo flux can be estimated from
\beq
\Nhalo = \frac{\Nhalo(\theta_1 < \thetah < \theta_2)}{g(\theta_2)-g(\theta_1)}
~~~;
\eeq
the function $g(\thetah)$ depends, of course, on assumptions concerning both
the grain model and the distribution of dust along the sightline.

\begin{figure}[h]
\centerline{\epsfig{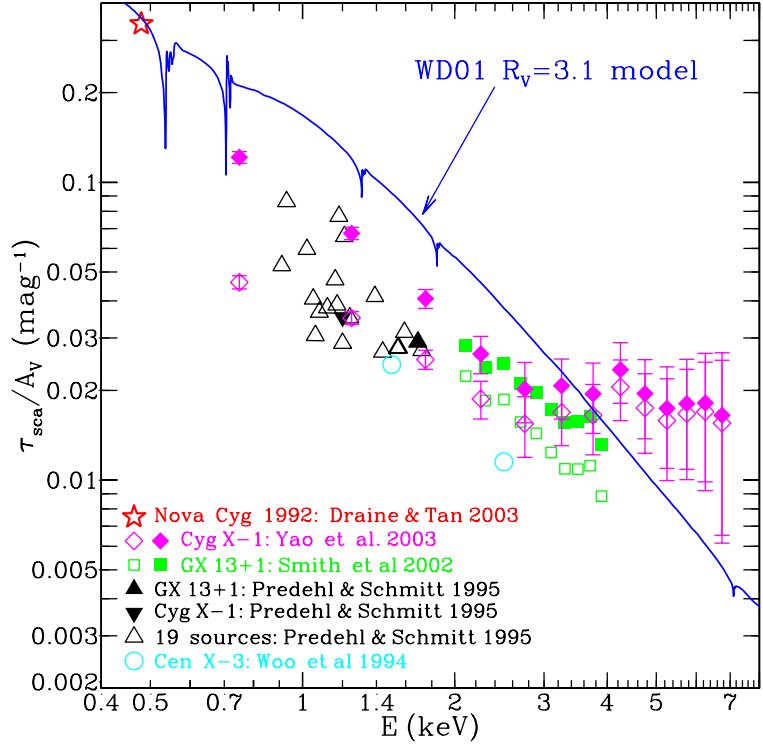}}
\caption{\footnotesize
	\label{fig:tau_s/A_V}
	Scattering optical depth $\tausca$ per unit visual
	extinction $A_V$ as calculated for the WD01 model (solid line) and
	as estimated from various observations (see text).
	\Chandra\ imaging of GX~13+1 at $50\arcsec<\theta_s<600\arcsec$ 
	(Smith, Edgar, \& Shafer 2002) has been corrected
	for photons interior to $50\arcsec$ assuming
	uniformly distributed dust (open squares) or dust with
	density proportional to distance
	(filled squares).
	\Chandra\ observations of
	Cyg X-1 at $\theta_s < 120\arcsec$ (Yao et al.\ 2003) are
	uncorrected for photons beyond $120\arcsec$ (open diamonds) or
	corrected assuming uniform dust (filled diamonds).
	See text.
	}
\end{figure}

\subsection{Cen X-3}

Woo et al.\ (1994) used the \ASCA\ X-ray observatory
to measure the X-ray halo toward the
massive X-ray binary Cen X-3
($A_V\approx 4.3$)
as a function of orbital phase, at 1.5 and $2.5\keV$.  
Their results for $\tausca/A_V$
are shown in Figure \ref{fig:tau_s/A_V}.  
Also shown is the ratio $\tau_{\rm sca}/A_V$ estimated for the
WD01 grain model.
Woo et al.'s values of $\tausca/A_V$ at 1.5 and 2.5~keV are
a factor $\sim$3.5 below the prediction of the WD01 grain model.
Can we understand this?

At 1.5~keV,
the \ASCA\ point spread function was such that even when the point source
component was at a minimum,
the halo intensity exceeded the point source 
intensity only
for $\thetah > 200\arcsec$; when the point source was at maximum the
halo intensity exceeded the point source intensity only for 
$\thetah \gtsim 450\arcsec$.
For uniformly-distributed dust and $E=1.5\keV$, we expect a
median halo angle
$\theta_{h,m}\approx 100\arcsec$ (see Table \ref{tab:thetah_values});
it is therefore clear that
the \ASCA\ observations were insensitive to most of the scattered photons.
It therefore seems 
plausible that the scattered flux may have been underestimated
by factors of 2-4 due to the dominance of the point source profile at
halo angles $\theta \ltsim 200-400\arcsec$.
The determination of $\tausca/A_V$ by Woo et al.\ should be treated as
a lower bound rather than a measurement.

\subsection{\ROSAT\ Observations}

\ROSAT\ observed both unscattered and scattered X-rays from Nova Cygni 1992
at a number of epochs (Krautter et al.\ 1996).
The most recent reanalysis of the \ROSAT\ data 
found 
$\tau_{\rm sca} = 0.211\pm0.006$ at the median photon energy
$\sim$480~eV (Draine \& Tan 2003).
Adopting $E(B-V)=0.19$ estimated from observations of H$\alpha$/H$\beta$
(Barger et al.\ 1993; Mathis et al.\ 1995), we obtain
$\tau_{\rm sca}/A_V=0.36$.
As seen in Figure \ref{fig:tau_s/A_V}, this value is in excellent agreement
with the WD01 grain model.
Draine \& Tan have also carried out detailed modelling, and conclude that
the observed X-ray halo profiles at 9 different epochs 
are in good agreement with the WD01 dust model.

Predehl \& Schmitt (1995, hereafter PS95) used \ROSAT\
to estimate
$\fhalo$ for
Cyg~X-1, GX~13+1, and 19 other
galactic sources for which $A_V$ was also available.
In each case, PS95 have fitted the observations with
the \ROSAT\ psf plus a theoretical dust model,
and used this to estimate the total number of scattered photons.
Figure \ref{fig:tau_s/A_V} shows the resulting
$\tausca/A_V$ 
versus
the average photon energy for each source.
The estimates of $\tausca/A_V$ inferred from the
PS95 observations are generally a factor $\sim2-4$ below the
WD01 grain model.
If the PS95 values of $\fhalo$ are accurate, this
would indicate a serious problem with the WD01 grain model.

However, for the 21 sources, the halo angle at which the intensities of
the fitted halo and psf were equal was $70\arcsec$ or larger; 
for 50\% of the sources this angle was $130\arcsec$ or larger.
At a typical energy of $\sim$1.2~keV, the median scattering angle is
$\thetahm\approx130\arcsec$ for uniformly-distributed dust
(see Table \ref{tab:thetah_values}).
Estimates of $\fhalo$ therefore
rely heavily on the dust model to
separate the halo from the point source
at small halo angles.
The modelling by PS95
employed power-law grain size distributions
$dn/da \propto a^{-q}$ for $a<a_{\rm max}$.
For the 21 sources, the median $q$ value was 4.0
and the median
value of $a_{\rm max}$ was $0.18\micron$.
Power-law grain size distributions with $q\approx 3.5$ and
$a_{\rm max}\approx 0.25\micron$ (Mathis, Rumpl, \& Nordsieck 1977;
Draine \& Lee 1984) provide a good fit
to the interstellar extinction, but
the size distributions adopted by PS95 -- with
generally steeper power laws and smaller values of $a_{\rm max}$ --
had insufficient mass
in large grains\footnote{%
	The median $q=4.0$ corresponds to equal mass per logarithmic
	interval; for $q\geq 4.0$ the grain mass diverges at small
	radii unless a lower cutoff is imposed.}
and do not provide a good fit to the extinction law.
The scattering at small halo angles is dominated by the larger grains,
so the PS95 model-fitting may have systematically
underestimated the actual halo intensity at small halo angles.

Given likely uncertainties in the psf and the model-fitting, it seems
likely that the PS95 values of $\tausca/A_V$ are
systematically low, perhaps by factors as large as 2-4.
We now examine two particular sightlines.

\subsection{GX~13+1}

GX~13+1 is a low-mass X-ray binary system
at $l=13.5^\circ$, $b=0.10^\circ$, and an estimated distance $D\approx7\kpc$,
corresponding to a distance above the plane $D\sin b\approx 12\pc$.
Garcia et al.\ (1992) estimated $A_V\ltsim 14.4$.
X-ray spectroscopy
gives $N_{\rm H}=2.9\pm0.1\times10^{22}\cm^{-2}$
(Ueda et al.\ 2001), corresponding to
$A_V=15.5$ for the standard conversion 
$N_{\rm H}/A_V=1.87\times10^{21}\cm^{-2}$ (Bohlin, Savage, \& Drake 1978).
We adopt $A_V=14$.
The mean photon energy for the \ROSAT\ observations is $E=1.69\keV$
(PS95).

PS95 find a halo fraction $\fhalo=0.335$ for
GX~13+1, but (as discussed above) this may be an underestimate:
\begin{itemize}
\item
For GX~13+1, PS95 estimate that the halo and psf have equal
intensities at $\sim$75\arcsec\ (see their Fig.\ 10).\footnote{
	The \ROSAT\ psf fit given by Boese (2000) has 
	90\%-enclosed-power radii of $32\arcsec$ at $0.5\keV$, 
	$25.5\arcsec$ at $1\keV$, $35\arcsec$ at $1.5\keV$,
	$52\arcsec$ at $1.69\keV$,
	and $95\arcsec$ at $2.0\keV$.
	It is not clear how close the actual psf is to the
	fit given by Boese.
	}
For uniformly-distributed dust and $E=1.69\keV$, 
the median halo angle $\thetahm\approx91\arcsec$, but on this
sightline the dust density may be enhanced closer to the source,
leading to a reduction in $\thetahm$;
for a linear gradient with $\beta=1$, $\thetahm\approx58\arcsec$.
Thus it appears possible that $\gtsim50\%$ of the scattered photons may have
been misattributed to the psf.\footnote{%
	The \ROSAT\ imaging extends to $\thetah\approx 2000\arcsec$, 
	but for GX~13+1 the background is estimated to exceed the halo 
	intensity for $\thetah \gtsim 900\arcsec$;
	determination
	of the background is itself difficult, and underestimation of
	$\Nhalo$ due to background oversubtraction at $\theta\gtsim 600\arcsec$
	is an additional possibility.  However, since 90\% of the scattering
	at 1.69~keV is at scattering angles 
	$\ltsim1080\arcsec/1.69=640\arcsec$,
	underestimation of the halo intensity at $\thetah>600\arcsec$ 
	would have only	a small effect on $\Nhalo$.
}
\item
For GX~13+1, PS95 used a dust model with $dn/da\propto a^{-3.8}$
and $a_{\rm max}=0.18\micron$; as discussed above, this
underestimates the abundances
of $a\gtsim 0.1\micron$ grains, and therefore underestimates the 
contribution of scattering at small
halo angles.
\end{itemize}
It therefore appears that PS95 could have underestimated 
$\Nhalo$.
While difficult to quantify, it seems possible that
the true value of $\fhalo$ might be as large as
0.65 (the value predicted by the WD01 grain model for $E=1.69\keV$ and
$A_V=14$).

The X-ray halo around GX~13+1 has recently been observed by the \Chandra\
X-ray telescope
(Smith, Edgar, \& Schafer 2002) at energies between 2.1 and $3.9\keV$,
with energy resolution $\sim 0.2\keV$.
Phenomena referred to as ``pileup'' and ``grade migration'' in the
detection system affect the \Chandra\ ACIS images as far as 
$50\arcsec$ from the source.
Smith et al.\ also discuss the current 
uncertainties concerning the \Chandra\
psf at angles $> 50\arcsec$.
Using a preliminary psf based on observations of Her X-1,
they estimate what they refer to as ``total observed halo fraction'' 
$I(E)$
by integrating the psf-subtracted and background-subtracted
count rates from $50\arcsec$ to $600\arcsec$,
and dividing by the estimated psf count rate 
in the absence
of saturation effects.
Thus
\beq
I(E) = \frac{(1-e^{-\tausca})[g(600\arcsec)-g(50\arcsec)]}
{e^{-\tausca}}
~~~.
\eeq
If we assume a model for the dust distribution, we can use
$g(\thetah)$ to obtain
\beq
\tausca=\ln\left[1+\frac{I}{[g(600\arcsec)-g(50\arcsec)]}\right]
~~~,
\eeq
where for a uniform gradient 
$g(\thetah)$ is given by eq.\ (\ref{eq:g(theta,beta)}).
In Figure \ref{fig:tau_s/A_V} we show the values of $\tausca/A_V$ obtained
from the observed $I$, with $g(\thetah)$ calculated for the WD01 grain model
assuming 
(a) uniformly-distributed dust ($\beta=0$) and 
(b) dust with a density gradient $\beta=1$.
Given the location of this source, we expect the dust density to be increasing
toward the source, so the $\beta=1$ model is reasonable.
For $\beta=1$, the inferred $\tausca/A_V$ is a factor $\sim$1.5 below the
WD01 model for 2.4--3.4~keV, and a factor 1.1--1.3 below the WD01
model at 3.4--4.0~keV.
These observations suggest that the WD01 model may overestimate $\tausca/A_V$,
although it should be kept in mind that these results required substantial
corrections for unobserved halo interior to $50\arcsec$.

The best way to use the information in the oberved scattered halo is to
try to reproduce the observed radial profile of the scattered halo using
a dust model, and
Smith et al.\ tested various grain models in this way.
For dust distributed uniformly between source
and observer, for the WD01 grain model they find a best-fit gas column 
$N_{\rm H}=1.65\times 10^{22}\cm^{-2}$, significantly
smaller than the value $2.9\pm0.1\times10^{22}\cm^{-2}$ estimated
from \ASCA\ observations (Ueda et al.\ 2001).
Note, of course, that additional dust could be located at $x > 0.75$ without
appreciably affecting the observed $I(\theta_h>50\arcsec)$, 
since the additional
halo contribution would be mainly 
below the $50\arcsec$ lower cutoff.\footnote{%
	$\theta_{h,m}=43\arcsec$ for $E=2.1\keV$ and
	dust at $x_d=0.75$.
	}

\subsection{Cygnus X-1}

Cygnus X-1 consists of an O star primary with a black
hole companion (see Tanaka \& Lewin 1995),
located at $l=71.33^\circ$, $b=3.07^\circ$, and an estimated distance
$D=2.5\pm0.4\kpc$ (Bregman et al.\ 1973; Ninkov et al.\ 1986), 
placing it at a height
$D\sin b\approx 130\pm20\pc$ above the plane.
The O9.7~Iab primary is reddened by $E(B-V)=1.12$ (Bregman et al.\ 1973),
corresponding to $A_V=3.5$; this is consistent with 
$N_{\rm H}=6.2\times10^{21}\cm^{-2}$ from
X-ray absorption spectroscopy (Schulz et al.\ 2002).
Based on studies of reddening vs. distance for stars 
within $50\arcmin$ of Cyg~X-1 
(Bregman et al.\ 1973; Margon et al.\ 1973) it appears
that the dust is distributed approximately uniformly along the
sightline.

PS95 observed Cygnus X-1 with \ROSAT, and found
$\fhalo=0.116$ at 1.2~keV.
The psf and scattered
halo intensity were estimated to be equal at $\theta\approx110\arcsec$.
The PS95 result $\tausca/A_V=.037\,{\rm mag}^{-1}$ at $1.2\keV$ is a factor 3
below the WD01 model (see Fig.\ \ref{fig:tau_s/A_V}).

\begin{figure}[h]
\centerline{\epsfig{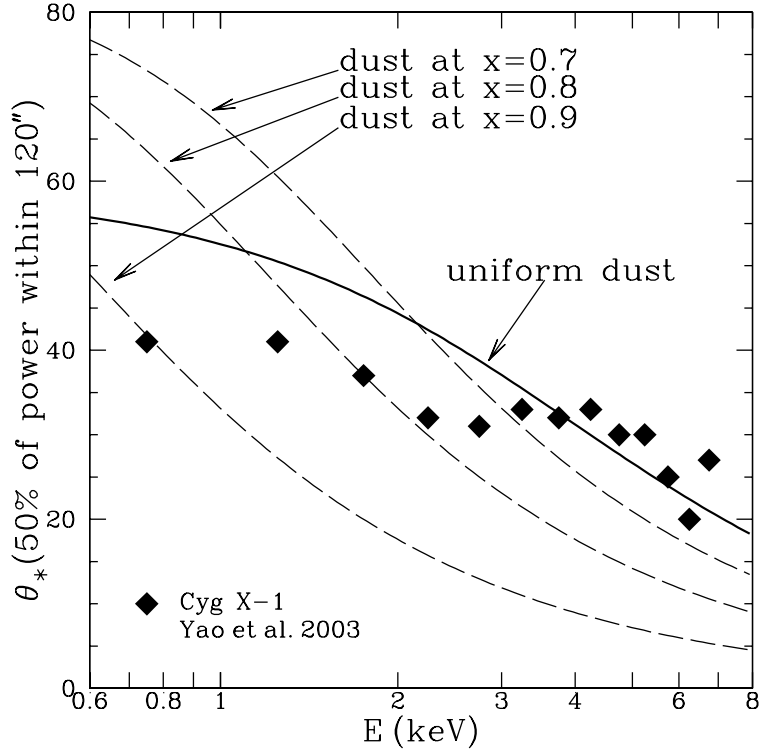}}
\caption{\footnotesize
	\label{fig:yaorad}
	Radius $\theta_*$ enclosing 50\% of the halo counts within $120\arcsec$
	of the point source.
	Data points are observations of Cyg~X-1 by Yao et al.\ (2003).
	}
\end{figure}

Yao et al.\ have recently used \Chandra\ observations of Cygnus X-1
to infer the scattered halo, using a technique designed to minimize the
effects of ``pileup'', allowing the excellent angular resolution of 
\Chandra\
to be used to observe at small halo angles.
Yao et al.\ neglected halo angles $\theta > 120\arcsec$, but were able to
measure the halo as close as $1\arcsec$ from the point source.
Their fractional halo intensity $\FHI$ is the ratio of the halo counts
interior to 120$\arcsec$ divided by the counts from the psf plus the
halo interior to 120$\arcsec$, and is related to $\tausca$ by
\beq
\FHI=
\frac{(1-e^{-\tausca})g(120\arcsec)}
{e^{-\tausca}+(1-e^{-\tausca})g(120\arcsec)}
~~~.
\eeq
Yao et al.\ also report the radius $\theta_*$ containing
50\% of the halo counts within $120\arcsec$ of the source, i.e.,
$g(\theta_*)=0.5g(120\arcsec)$.
In Figure \ref{fig:yaorad} we show the variation of $\theta_*$ with
$E$ calculated for: (1) uniform dust;
(2) thin sheet at $x_d=0.7$; (3) thin sheet at $x_d=0.8$;
(4) thin sheet at $x_d=0.9$.
The single sheet models clearly do not reproduce the observations.
The uniform dust model is in approximate overall agreement with
the distribution of halo counts within 120$\arcsec$, although it does
not reproduce the concentration of the halo at $E < 3\keV$
found by Yao et al.

Adopting the uniform dust model, we
can now use the WD01 model to correct for halo counts at $\thetah>120\arcsec$:
$\tausca = \ln \{1+\FHI/[(1-\FHI)g(120\arcsec)]\}$, with the results
plotted in Fig.\ \ref{fig:tau_s/A_V}.
At $E<2.5\keV$, where the estimated fractional uncertainties
are smallest, the values of $\tausca/A_V$ found from the Yao et al.\ 
results fall a factor $\sim1.5-2$ below the predictions of the WD01 model.
The reason for this is not apparent.
Perhaps the WD01 model has overestimated $\tausca/A_V$; alternatively,
perhaps the background has been overestimated -- background oversubtraction
would be consistent with the surprising concentration of the halo
seen at $E<2.5\keV$ in Fig.\ \ref{fig:yaorad}.

For 3--4$\keV$,
the inferred values of $\tausca/A_V$, and the measured halo half-light radii
in Fig.\ \ref{fig:yaorad}, are consistent with the
WD01 dust model.
However, at $5-7\keV$ the values of $\tausca/A_V$ measured by Yao et al.\
exceed the predictions of the WD01 model, although the error bars
are now large.
It is difficult to envision a dust model which could have such
a large value of $\tausca/A_V$ at these energies, so we suspect that
the observations are affected by some systematic error --
perhaps the wings of the 
\Chandra\ psf at these energies may have been underestimated,
or the novel 
method employed by Woo et al to reconstruct the radial profile may be
prone to systematic errors which are not fully understood.

\subsection{Discussion}

We have reviewed a number of measurements of dust scattering halos,
and compared the predictions of the WD01 grain model to the
values of $\tausca/A_V$ estimated from these observations.
The results, shown in Fig.\ \ref{fig:tau_s/A_V}, are somewhat equivocal.
Two studies (Woo et al.\ 1994; Predehl \& Schmitt 1995) find values
of $\tausca/A_V$ much lower than expected for the WD01 dust model,
but we give arguments why these observations might have 
underestimated $\tausca$.

Draine \& Tan (2003) have quantitatively
modelled the X-ray halo around 
Nova Cygni 1992 (typical photon energy $\sim$0.48~keV) 
using the WD01 grain model.
The halo intensity observed by \ROSAT\ 
can be reproduced using a dust column density
which agrees with the reddening inferred from the observed
H$\alpha$/H$\beta$ intensity ratio. 

\Chandra\ observations of GX~13+1 by Smith et al.\ (2002), after
correcting for missed halo counts using the constant dust gradient
model with $\beta=1$, give $\tausca/A_V$ within a factor $\sim$1.5 of the
WD01 model at 2~keV, and in agreement at 3.5--4~keV.
The corrections are somewhat sensitive to
the (uncertain) dust density distribution for this case, since
the halo intensity was not measured at $\thetah<50\arcsec$.

The recent \Chandra\ measurement  of the halo around Cyg~X-1 
(Yao et al.\ 2003) implies values of $\tausca/A_V$ which are a factor
$\sim$1.5--2 smaller than expected for the WD01 model
at $E<2.5\keV$.
At 3--4~keV the Yao et al.\ results are in excellent agreement with the
WD01 model in terms of both $\tausca/A_V$ and the half-light radius
of the halo.
For 5--7~keV the values of $\tausca/A_V$ found by
Yao et al.\ exceed the WD01 model by factors of 2--3, although the
estimated errors are also large.

Taken together, 
the \Chandra\ observations of GX~13+1 (Smith et al.\ 2002)
and of Cyg~X-1 (Yao et al.\ 2003) suggest that the WD01 model may
have overestimated $\tausca/A_V$ by a factor $\sim$1.5 between 1 and 2.5~keV.
However, the Yao et al observations of Cyg X-1 at 3--4~keV give both
a halo concentration and $\tausca/A_V$ in good agreement with the WD01
model, and the WD01 model is also consistent with the Nova Cygni observations
at $\sim$0.5~keV (Draine \& Tan 2003).
At this time we can conclude only
that the WD01 model appears to give $\tausca/A_V$ in agreement
with observations to within a factor $\sim$1.5, but the existing
observations do not permit a more precise statement.

It is hoped that future observations by \Chandra\ or {\it XMM}
will be able to carry
out high-signal-to-noise observations of X-ray scattering halos
on sightlines where the dust distribution and reddening are well-determined.
An optimal situation would be to use a source which is known
to be distant compared to the dust doing the scattering, so that we can
assume that $x \ltsim 0.2$ in eq.\ (\ref{eq:thetah_thetas}).
An extragalactic X-ray point source (AGN or quasar)
would be ideal for this purpose.

\section{Summary\label{sec:summary}}

The following are the principal results of this work:
\begin{enumerate}
\item Dielectric functions for graphite and MgFeSiO$_4$ have been
constructed which are continuous from submm to hard X-rays, 
obey the Kramers-Kronig relations, and satisfy the
oscillator strength sum rule.

\item Absorption, scattering, and extinction have been calculated for
the WD01 grain model at X-ray energies.
The calculated absorption edge structure appears to be consistent with
recent spectroscopy by the {\it Chandra X-Ray Observatory}, and can
be further tested by future observations.

\item Differential scattering cross sections are presented for the
Milky Way dust model at X-ray energies.
These can be used for modelling X-ray scattering halos, and are
available at http://www.astro.princeton.edu/$\sim$draine .

\item The median scattering angle $\thetasm(E)$ is
given, as well as the scattering angles $\thetasten$ and $\thetasninety$
for 10\% and
90\% enclosed power.  These can be used to assess the sensitivity of
imaging observations for determination of the flux of halo photons.

\item Simple analytic functions
provide a good approximation to the differential scattering cross section
for dust  (eq.\ \ref{eq:dsigdom_approx},\ref{eq:sigma<theta}).

\item We provide analytic approximations to $g(\thetah)$,
the fraction
of the halo counts interior to $\thetah$, for dust in a single sheet
(eq.\ \ref{eq:g_sheet}) and
uniformly-distributed dust or dust with a density gradient
(eq.\ \ref{eq:g(theta,beta)}).

\item The total scattering cross section calculated for the WD01 grain
model is compared with observations of X-ray halos by {\it ASCA},
\ROSAT, and \Chandra\ (see Fig.\ \ref{fig:tau_s/A_V}).  The results
are somewhat equivocal, and in some cases depend on corrections which
are sensitive to the spatial distribution of the dust.  \ROSAT\
observations of Nova Cygni 1992 at $\sim$0.5~keV (Draine \& Tan 2003),
and \Chandra\ observations at 3--4~keV of Cyg X-1 (Yao et al.\ 2003)
are in good agreement with the WD01 dust model, although the 2--3.4~keV
observations of GX~13+1,
and 1--3~keV observations of Cyg~X-1, 
give $\tausca/A_V$ about a factor $\sim$1.5
below the WD01 model.  At this time it is possible to conclude from
the observations that $\tausca/A_V$ given by the WD01 model is
accurate to within a factor $\sim$1.5, but a more precise statement is
not possible.

\item
Analysis of the angular structure of X-ray halos around Galactic sources
will generally be compromised by uncertainties concerning the location of
the dust responsible for the scattering.
The ideal observation is to observe an extragalactic point source, in
which case the scattering dust is all at $x\approx0$.
It is hoped that such observations will be carried out by \Chandra\ 
or {\it XMM}
for bright extragalactic sources located behind sufficient Galactic
dust, thereby providing a definitive test of this and other dust models.

\end{enumerate}
\acknowledgements

I am grateful to Yangsen Yao for making the Cyg X-1 results available
in advance of publication.
I thank
Aigen Li, Randall Smith, and
Jonathan Tan for valuable comments, an
anonymous referee for helpful suggestions,
and Robert Lupton for making available the SM software package.
This work was supported in part by
NSF grant AST-9988126.


\end{document}